\documentclass[twocolumn,aps,prb,showpacs]{revtex4}

\usepackage{graphicx}
\usepackage[usenames,dvipsnames]{xcolor}
\usepackage{hyperref}
\definecolor{darkblue}{rgb}{0,0,.75}
\hypersetup{colorlinks=true, breaklinks=true,
linkcolor=darkblue, menucolor=darkblue,
urlcolor=darkblue,citecolor=darkblue}

\usepackage{amsmath}
\usepackage{amssymb}

\newcommand{\bk}{\mathbf{k}}
\newcommand{\bp}{\mathbf{p}}
\newcommand{\br}{\mathbf{r}}
\newcommand{\bx}{\mathbf{x}}
\newcommand{\ba}{\mathbf{a}}

\newcommand{\bq}{\mathbf{q}}

\newcommand{\delp}{{\Delta\mathbf{p}}}

\newcommand{\taue}{\tau_\text{e}}

\newcommand{\deleta}{{\Delta\eta}}

\newcommand{\trt}{T} 
\newcommand{\delt}{\trt}
\newcommand{\rmT}{\mathrm{T}} 
\newcommand{\T}{\mathrm{T}} 
\newcommand{\K}{\mathrm{K}} 

\newcommand{\trsp}{\mathrm{t}}

\begin{document}

\title{Echo spectroscopy of Anderson localization} 
 
\author{T.~Micklitz$^1$, C.~A.~M\"uller$^{2,3}$, and A.~Altland$^4$}

\affiliation{$^1$Centro Brasileiro de Pesquisas F\'isicas, Rua Xavier Sigaud 150, 22290-180, Rio de Janeiro, Brazil \\
$^2$Fachbereich Physik, Universit\"at Konstanz, 78457 Konstanz, Germany\\
$^3$Institut nonlin\'eaire de Nice, Universit\'e Nice--Sophia Antipolis, CNRS, 06560 Valbonne, France\\  
$^4$Institut f\"ur Theoretische Physik, Universit\"at zu K\"oln, Z\"ulpicher Str. 77, 50937 K\"oln, Germany}

\date{\today}


\begin{abstract}

We propose a conceptually new framework to study the onset of Anderson
localization in disordered systems. The idea is to expose waves propagating in
a random scattering environment to a sequence of short dephasing pulses. The
system responds through coherence peaks forming at specific echo times, 
each echo representing a particular process of quantum interference. We suggest a concrete
realization for cold gases, where quantum interferences are 
observed in the momentum distribution of matter waves in a
laser speckle potential,  and discuss in detail corresponding echoes in momentum space for 
sequences of one and two dephasing pulses.  
Our proposal defines a challenging, but arguably realistic framework promising to yield unprecedented 
insight into the mechanisms of Anderson localization.
\end{abstract}

\maketitle

\section{Introduction}

Coherent chaotic scattering is a defining feature of disordered
quantum systems. Its manifestations range from coherence peaks in scattering
cross sections over weak localization and quantum fluctuation phenomena in
metals, to strong Anderson localization \cite{50yearsAL}. 
Phenomena of this
type have been observed 
with light~\cite{ALlight1,ALlight2} or microwaves~\cite{ALmicrowaves},
in electronic conductors~\cite{AL}, with cold atomic
gases~\cite{Chabe2008,Billy2008,Roati2008,Kondov2011,AL3DPalaiseau,AL3DFlorence}, photonic
crystals~\cite{ALpc1,ALpc2}, and classical waves \cite{ALothers}. 
Semiclassically,
quantum coherence is understood in terms of the interference of Feynman path
amplitudes. Quantum effects arise when classically distinct amplitudes
interfere to yield non-classical contributions to physical observables, see
Fig.~\ref{fig1}. For instance, coherent backscattering (CBS) and weak
localization \cite{akkermans2006} are due to the interference of mutually time
reversed paths. Similarly, coherent forward scattering is caused by the
concatenation of two such processes, or again by the interference of two self
retracing loops traversed in different order \cite{Karpiuk2012,Micklitz2014,Lee2014},
etc. Quantum coherent contributions are often discriminated from classical
background contributions by their strong sensitivity to dephasing and
decoherence. 
However, other than suppressing
coherence, generic sources of decoherence -- external magnetic fields, 
AC electromagnetic radiation, etc. -- do not provide much insight into the
mechanisms of quantum interference in disordered media. Furthermore,
decoherence often acts as a source of heating (it certainly does so on the
temperature scales relevant to cold atomic gases) and leads to an unwelcome
nonequilibrium shake-up of the system.

In this paper, 
we suggest an alternative protocol for probing quantum coherence.
Its advantage is that it offers much more specific
information and at the same time is less intrusive than persistent
external irradiation. The idea is to expose the quantum system to a source of dephasing
only at specific `signal times',
$t_1,t_2,\dots$. The system then responds to this perturbation at
`echo times' $\tau_1,\tau_2,\dots$, which are in well-defined correspondence to the
signal times. Each of these echoes corresponds to a specific mechanism of 
quantum-coherent scattering. For example, an echo at time $2t_1$ after a
dephasing pulse applied at time $t_1$ is a tell-tale signature of
the CBS effect (see Fig.~\ref{fig1} below). 
Likewise, an echo observed at time $2(t_2-t_1)$ in response to \textit{two} pulses at
$t_1$ and $t_2>2t_1$ identifies a contribution to forward scattering coherence, 
etc. The observation of a temporal echo pattern thus realizes a highly resolved probe of quantum
coherence in random scattering media.

The rest of the paper is organized as follows. 
In Section \ref{sec:semiclassical_approach_to_coherence_echoes}
we introduce the Feynman-path approach to coherence echoes and discuss 
real-space echo signals up to two-loop order.
Section~\ref{MSE} discusses the first-order coherence echo in momentum space, while details about 
the second-order momentum-space signal are relegated to Appendix~\ref{app:Det}. Section~\ref{sec:FT} 
contains the systematic derivation of all results within a field-theoretical formalism.
In the concluding Section \ref{Conclusion}
we suggest an experimental realization of echo spectroscopy with cold
atoms. Further details on diffusion mode calculations 
are contained in Appendix \ref{app:Diff}.

\section{Feynman path approach to coherence echoes}
\label{sec:semiclassical_approach_to_coherence_echoes}

We consider a $d$-dimensional system of non-interacting 
quantum particles moving in a random potential and described by the Hamiltonian 
\begin{align}
\label{h}
\hat{H}={\hat{\bold{p}}^2\over 2m} + V(\hat{\br}). 
\end{align} 
The random potential $V$ is assumed to be an uncorrelated Gaussian
process with covariance  
$\langle V(\br) V(\br') \rangle
= {1\over 2\pi\nu\tau}\delta(\br-\br')$, 
where $\nu$ is the density of states per volume and $\tau$ the elastic scattering time. 
Central for our discussion is the retarded quantum correlation function
\begin{align}
	\label{eq:Xdef}
	X\equiv  \left\langle \hat O_\bx(t)\hat O_{\bx'}(0)\right\rangle,
\end{align}
where the brackets stand for an average over quantum and disorder distributions,  
and $\hat{O}_\bx=|\bx\rangle\langle \bx|$ is 
a projector onto a squeezed state defined by $\left\langle
\br'|\bx\right\rangle=\frac{1}{(2\pi)^{d/4}}\frac{1}{(\Delta
r)^{d/2}}\exp\left(-\frac{(\br'-\br)^2}{(2\Delta r)^2}+\frac{i}{\hbar}\bp\cdot \br'\right)$.
The scale $\Delta r$ sets the spatial resolution of the operator, 
and  $\bx=(\br,\bp)$ is a phase space vector comprising real space ($\br$)
and momentum space ($\bp$) coordinates.  
In the
limit of infinitely sharp resolution $\hat O_\bx\stackrel{\Delta
r\to 0}\longrightarrow |\br\rangle\langle \br|$ projects onto  real-space
coordinates, and the correlation function \eqref{eq:Xdef} may serve, 
e.g., as a building block for a point-contact transport observable. In the
opposite limit  $\hat O_\bx\stackrel{\Delta r\to \infty}\longrightarrow
|\bp\rangle\langle \bp|$ projects onto momentum coordinates, and the correlation
function relates to the cross section for the scattering process $\bp \to \bp'$.
Intermediate values of $\Delta r$ probe transitions between
coherent-state-like wave packets of minimal quantum uncertainty centered
around $\bx$.

\begin{figure}[b]
\centering
\includegraphics[width=8.5cm]{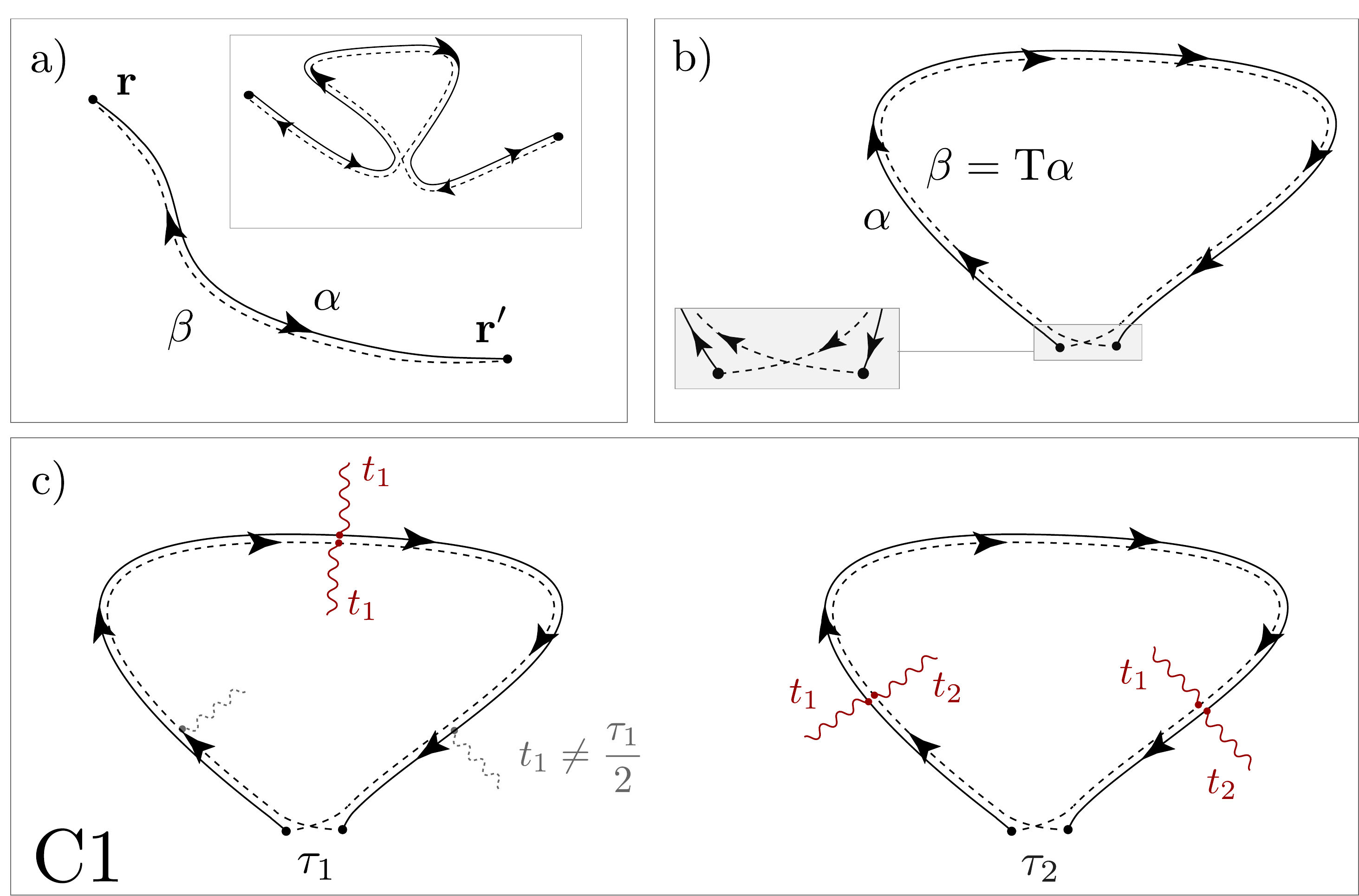}
\caption{\label{fig1}
Physical observables represented in terms of pairs of retarded (solid lines) and advanced (dashed lines) Feynman path amplitudes.
a) Copropagating Feynman paths, $\alpha=\beta$, yield the classical contribution to the two-point transition probability $\br\to
   \br'$. Inset: weak
   localization loop. b) Coherent contribution, $\beta=\mathrm{T}\alpha$ to return probability $\br\to
   \br$, where $\mathrm{T}\alpha$ is the time reverse of $\alpha$. c) Coherent backscattering contribution in the presence of 
   dephasing pulses (wiggly lines).  While a pulse at time $t_1$ (dashed wiggly lines) suppresses the phase coherence of generic loops, 
   it affects particle and hole amplitudes in synchronicity if the loop is traversed in time $\tau_1=2t_2$, where coherence is briefly restored. 
   Right: synchronicity condition for a
  bi-temporal pulse at times $t_{1,2}$ is realized at traversal time
  $\tau_2=t_1+t_2$, where a coherence signal is observed.}
\end{figure}

To introduce the concept of coherence echoes, we consider in this section the case $\Delta r=0$ 
of a space-local two point correlation function. Within a Feynman path approach the 
expectation value \eqref{eq:Xdef} then assumes the form~\cite{cmft}
\begin{align}\label{Xs}
	X=\sum_{\alpha,\beta}\,\left\langle e^{\frac{i}{\hbar}
	(S[\alpha]-S[\beta])}M_{\alpha\beta}\right\rangle,
\end{align}
where $\alpha,\beta$ are paths connecting  $\br$ and $\br'$ in time $t$,
$S[\alpha]$ is the corresponding classical action, $M_{\alpha\beta}$ is a
container symbol for matrix elements and semiclassical stability amplitudes, 
and brackets stand for an average over disorder configurations. 
The double sum is
dominated by path configurations of nearly identical action
$|S[\alpha]-S[\beta]|\lesssim \hbar$, all other contributions are
effectively averaged out by large  phase fluctuations. The set of
contributing paths includes $\alpha=\beta$ [Fig.~\ref{fig1}a)], which yields the classical, 
phase-insensitive approximation $X_0$ of the observable \eqref{Xs}. 
Quantum corrections arise when paths branch out and subsequently recombine to form a phase
coherent  correction [Fig.~\ref{fig1}a) inset]. One may think of the internal loop included in this process 
as a `self energy' modifying the classical propagation in terms of a loop returning to its point of origin [Fig.~\ref{fig1}b)]. 
It is these loop structures, with external `classical legs' detached that are probed by our present approach: 
coherence signals tested by echoes arise when the two observation points $\br \to
\br'$ approach each other [Fig.~\ref{fig1}b)]. The double sum is then given by an uninteresting classical contribution $\alpha=\beta$,
and an \textit{equally strong} quantum contribution $\beta=\mathrm{T}\alpha$,
where $\mathrm{T}\alpha$ is the time reversed of the path $\beta$, and which is equivalent to the above self energy correction. 

Consider now a single external
radiation pulse applied to the system at time $t_1>0$ [Fig.~\ref{fig1}c)].  At $t_1$ a particle
propagating along $\alpha$ is at coordinate $\mathbf{r}(t_1)$, while a
particle propagating along $\mathrm{T}\alpha$ is at $\mathbf{\br}(t-t_1)$,
where $t$ is the loop traversal time selected by the moment of observation.  In general,
these coordinates differ from each other, which means that the external pulse
affects the quantum phases carried by the two amplitudes in different ways---causing dephasing. 
However, if the traversal time is such that $t_1=t/2$, then $\mathbf{r}(t_1)=\mathbf{r}(t-t_1)$, 
and coherence is briefly regained. 
Another way of stating the same fact emphasizes the time reversal 
symmetry essential to the coherent backscattering signal: at time $t =2t_1$, time reversal $t\to 2t_1-t$
relative to the signal time $t_1$ is restored and the conditions for phase coherence apply. 
An observation of the system at
time $t=2t_1\equiv \tau_1$ probes path pairs of just this `resonant' length, which can be witnessed by 
the formation of a coherence peak in the observable $X$.

\subsection{Perturbed quantum diffusion}
\label{PerturbedQD}

To obtain a quantitative understanding of the echo signal, 
we consider a weakly disordered medium in which 
the paths entering individual segments of pair
propagation (the double lines in Fig.~\ref{fig1}) 
describe diffusion. For fixed initial and final coordinates $\br$ and $\br'$  and
propagation time $t$, the sum over all co-propagating paths is described by a classical diffusion
propagator $\Pi_\mathrm{D}(\br,\br';t)$, or `diffuson' for brevity \cite{akkermans2006}. 
The diffuson solves the diffusion equation
$(\partial_t - D\partial^2_\br)\Pi_\mathrm{D}(\br,\br';t)=\delta(\br-\br')\delta(t)$,
where $D=v^2\tau/d$  is the classical diffusion coefficient, $\tau$ the
elastic scattering time, and $v= |\bold{p}|/m$ the velocity of particles of
mass $m$. Likewise, the sum over all contributions to a segment $\br\to \br'$
of counter-propagating paths is described by the propagator 
$\Pi_\mathrm{C}(\br,\br';t)$, the so-called Cooperon mode, which in the
absence of dephasing obeys the same diffusion equation.

\begin{figure} 
\centering
\includegraphics[width=6.5cm]{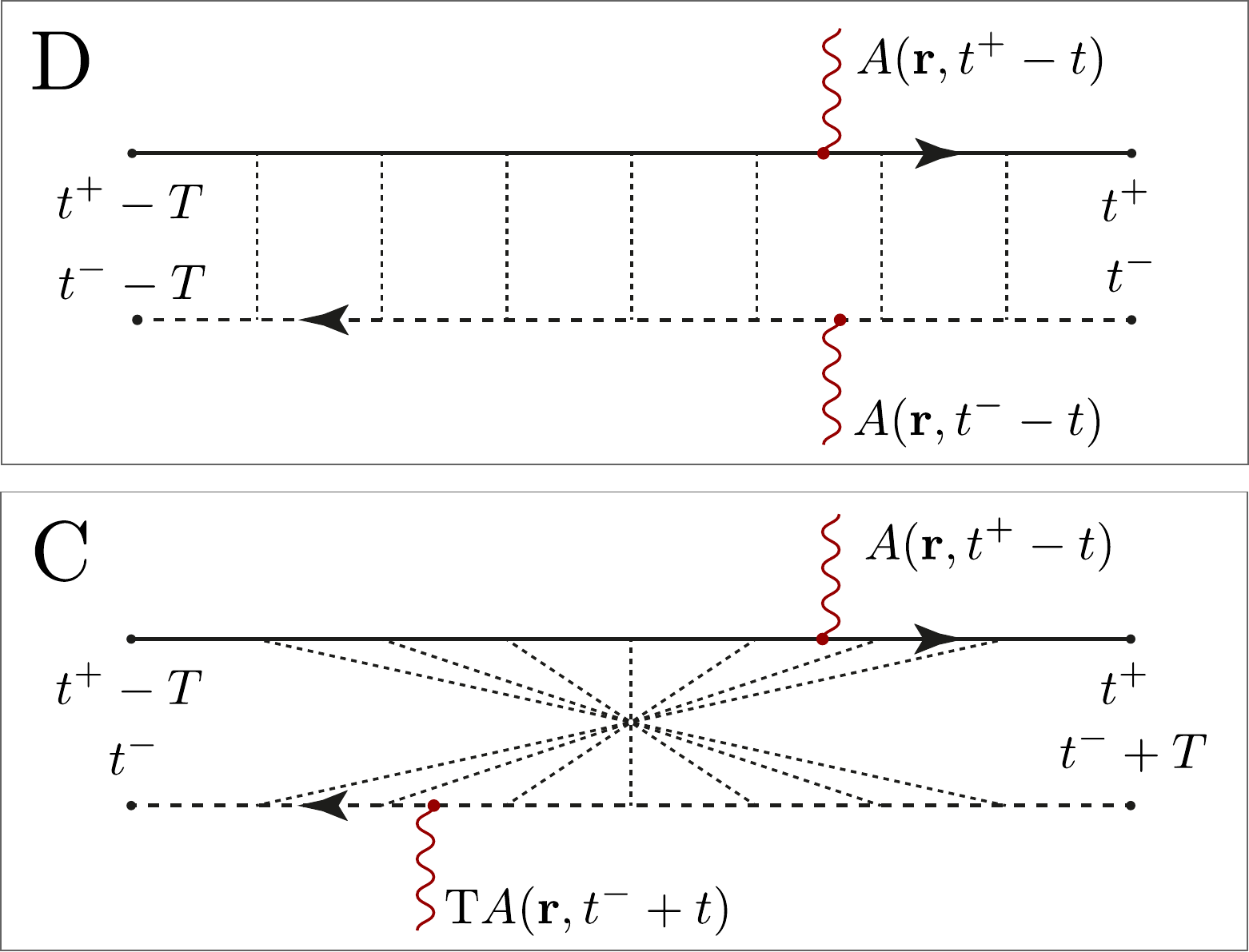}
\caption{\label{DiffusonCooperon}
Coupling of diffuson (D) and Cooperon (C) to an external
field $A=(\phi,\ba)$. 
The field acts at  
position $\br$ at the
passage time of particle (solid upper line) and hole (dashed lower line),
respectively. If these positions differ, dephasing occurs. 
In panel C, $\rmT A = \rmT(\phi,\ba) = (\phi,-\ba)$ 
indicates time reversal. 
}
\end{figure}

Let us now consider diffusive propagation in the presence of an external
source of radiation, represented by a four-potential $A=(\phi,\ba)$,
comprising a scalar and a vectorial component, $\phi=\phi(\br,t)$ and
$\ba=\ba(\br,t)$, resp. To account for the externally imposed time dependence
in a quantum diffusive process, we need to keep track of the traversal times
of the participating Feynman paths. The situation is illustrated in
Fig.~\ref{DiffusonCooperon}, where `D' is a diffuson mode comprising two
amplitudes starting at times $t^\pm -T$, resp., and ending at $t^\pm$. We denote generally 
by $T$ the time required to traverse the segment, and the dashed lines are
symbolic for the quantum scattering events causing diffusion. The wiggly lines
represent the action of the external field at time $t^\pm -t$. If 
the two paths are traversed simultaneously, $t^+=t^-$, the potential affects the upper
and lower line in the same way. In this case, the field does not 
destroy the mode, which is another way of saying that classical
diffusion is not affected by quantum decoherence. In the Cooperon process, `C', 
scattering paths are traversed in opposite order, as indicated by the `maximally
crossed' representation of scattering vertices. (Equivalently, one may flip the lower line, 
which leads to the un-crossed representation with co-oriented arrows employed in the 
rest of the figures). The sign change in the time reversed
potential $\rmT A = \rmT(\phi,\ba) = (\phi,-\ba)$ reflects the time reversal
symmetry breaking nature of external vector potentials. Likewise, a time-dependent scalar potential 
$\phi(t)$ will cause dephasing, unless an echo condition is met. 

The influence of the field on the diffusion modes can be
quantitatively described by diagrammatic perturbation theory \cite{EfrosPollack}. 
Under the assumption that the external field alters quantum phases but is sufficiently weak 
not to change the classical trajectories themselves, the perturbed 
diffuson and Cooperon modes (${\rm M}={\rm D,C}$)
are still governed by generalized diffusion equations 
\begin{align}
	\label{eq:DiffusionA}
	&\mathcal{D}_{\mathrm M} \Pi_\mathrm{M}(\br,\br';t^+,t^-,\trt)
	=\delta(\trt)\delta(\br-\br'), 
	\cr
&\quad \mathcal{D}_{\mathrm D/C} = \partial_{t^+}\pm\partial_{t^-} -i[\phi(\br,t^+) -
          \phi(\br,t^-)] \nonumber \\
	& \hspace{2cm} -
	D\left(\partial_{\br}+i [\ba(\br,t^+)\mp\ba(\br,t^-)]\right)^2,
\end{align}
in which the field $A= (\phi,\ba)$ enters through a covariant derivative.
For a given $A$, these `imaginary-time Schr{\"o}dinger equations' can be solved,
e.g., by path-integral techniques \cite{EfrosPollack,Feynman} (see Appendix~\ref{app:Diff}). 
We here
consider a situation without magnetic field, $\ba=0$, and a scalar potential
\begin{equation} 
\label{dephasingpot}
\hbar\phi(\br,t)=-  \br \cdot \Delta\mathbf{p} f(t). 
\end{equation} 
In the remainder of this section, we take $f(t) =\sum_{i=1}^N\delta(t-t_i)$ to represent a sequence of short pulses, 
each of which applies a homogeneous force that transfers a momentum $\delp$; the above weak field
assumption requires that each transfered momentum be much smaller than the particle momentum, $|\delp|\ll
p$. The dephasing pulses realize the ideal
form of effective $\delta$-kicks if they are shorter than the mean free
time $\tau$. Longer pulses are also admissible and provide full
echo contrast as long as they are symmetric around $t_i$.

\subsection{First-order echo signal}
\label{FirstOrder}

The first-order quantum
coherence contribution to the observable \eqref{eq:Xdef} involves two counter-propagating paths running 
synchronously between time $0$ and $t$, and thus  
has the time arguments $t^+=t,t^-=0,T=t$. 
 For times
$t<t_1$ before the first pulse,  the single Cooperon contribution
$X_{\rm C1}(t)=c/(D t)^{d/2}$ is just the classical probability of return
within time $t$, where $c$ is a numerical constant. 
Around the time $t=2t_1$, the signal is found to behave like 
\begin{align}
\label{singleCooperonecho}
\delta X_\text{C1}(t) 
&=X_{\rm C1}(t) e^{-  |t-2t_1|/\tau_\mathrm{e} }.
 \end{align} 
This describes a near instantaneous destruction of the coherence
 contribution by the pulse at $t_1$ followed by a revival at the echo time
 $\tau_1=2t_1$ over a width 
 \begin{equation} 
 \label{taue}
 \tau_\mathrm{e}=\hbar^2/D\Delta p^2.
 \end{equation} 
 The complete derivation of this signal, allowing also for a generalization to more general pulse profiles, 
 follows from the momentum-space results as described in Sec.~\ref{MSE} below. But the echo 
 profile \eqref{singleCooperonecho} can be readily understood by noting that the phases of the two
 amplitudes are affected as 
 \begin{equation} 
 \left\langle
 e^{i[\phi(\br(t_1))-\phi(\br(t-t_1))]}\right\rangle\simeq
 e^{-\frac{1}{2}\left\langle
 [\phi(\br(t_1))-\phi(\br(t-t_1))]^2\right\rangle},
 \end{equation} 
 where the angular
 brackets represent averaging over path configurations. Substituting the
 potential \eqref{dephasingpot} and noting that for a diffusive process $\left\langle
 [\br(t)-\br(s)]^2\right\rangle \sim D |t-s|$ one then obtains
 \eqref{singleCooperonecho}. Also, one sees that the characteristic echo time \eqref{taue} is determined by the time
 scale over which the phase mismatch between the two
 amplitudes reaches unity, $\left\langle (\Delta p\Delta x)^2\right\rangle\sim
 \Delta p^2 D \tau_\mathrm{e}= \hbar^2$.

The first-order coherence signal \eqref{singleCooperonecho} is suppressed directly after the C1 echo at time $\tau_1$. 
If, now,  a second pulse is applied at time
$t_2>\tau_1$, the coherence condition is met once more at $\tau_2 \equiv
t_1+t_2$, and another C1 echo will be observed [Fig.~\ref{fig1} c) second
diagram]. In addition to this signal, however, such a bi-temporal pulse gives
rise to further echoes, which probe more complex manifestations of quantum interference, to be discussed next.

\begin{figure}[b]
\centering
\includegraphics[width=8.5cm]{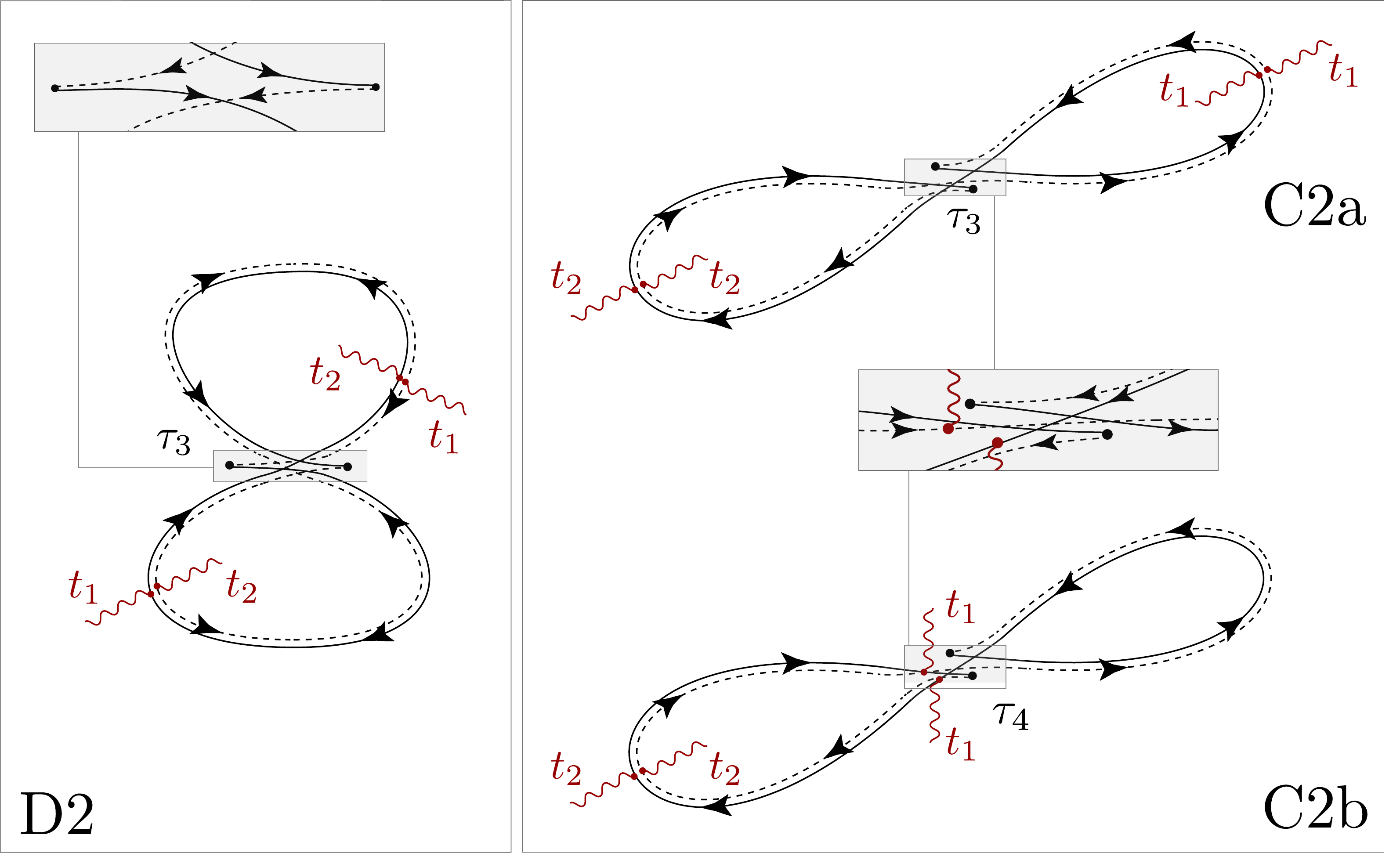}
\caption{\label{2ndOrderEchoReal}
Higher-order coherence contributions to the return amplitude probed by bi-temporal pulsing. 
The dephasing vertices shown in the inset of the right panel are only present in the C2b process. 
For the definition of the observation times $\tau_{3,4}$ and further discussion, see text.}
\end{figure}

\begin{figure*}  
\begin{center}
\includegraphics[width=16cm]{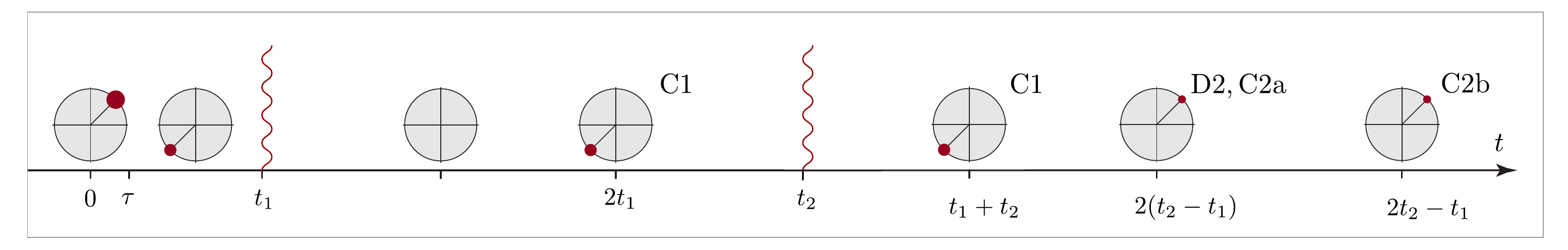}
\caption{\label{PeakChronology}
Chronology of quantum coherence echoes in the $k_x{-}k_y$ plane of momentum space. Echoes are indicated 
by red dots whose width/position hint at the signal strength/angular orientation on the elastic scattering manifold. 
From left to right: 
An initial state with well defined momentum yields the single-Cooperon (C1) backscattering peak after the transport 
time $\tau$. A first dephasing pulse at $t_1$
 supresses the C1 signal, which reappears at the first echo time $\tau_1=2t_1$.  
A second pulse at $t_2$ generates the bi-pulse C1 echo at $\tau_2=t_1+t_2$. 
Two-mode echoes appear in the forward scattering direction at $\tau_3=2(t_2-t_1)$ (D2, C2a) and $\tau_4=2t_2-t_1$ (C2b).}
\vspace{-.4cm}
\end{center}
\end{figure*}

\subsection{Probing higher-order quantum interference}
\label{HigherOrder}

We find that a double pulse selectively generates echo signals from two-loop contributions, as depicted in 
Fig.~\ref{2ndOrderEchoReal}. Consider, for example, the D2 coherence process that describes the interference of paths  along two loops which are
traversed in the same direction (no time reversal required!), but in different order. During its traversal of the first loop, the
particle is hit by the first pulse at time $t_1$. The particle then moves on into the
second loop, where it is hit by the second pulse at time $t_2$.  A straightforward assignment
of travel times to path segments shows that the hole amplitude (going through the loops in opposite order) 
will experience the pulses in synchronicity, i.e. at the same spatial path coordinates, 
provided the time of traversal for each loop be $t_2-t_1$.
In this case, the process becomes coherent, and an echo will be observed at $\tau_3\equiv 2(t_2-t_1)$.

A similar argument shows that at the same time $\tau_3$ the Cooperon process
C2a shown in Fig.~\ref{2ndOrderEchoReal}---consisting of two \emph{counter}-propagating
loops traversed in the same order---becomes phase coherent, too. For that
path configuration the coherence condition is satisfied at one more time 
$\tau_4\equiv 2t_2-t_1$ and this leads to one more echo C2b, also indicated in
Fig.~\ref{2ndOrderEchoReal}. 
Quantitative calculations below result in the two-loop echo contributions 
\begin{align}
\delta X_{\rm M}(t) 
=X_{\rm M}(t) e^{-|t-\tau_\mathrm{M}|/\tau_\mathrm{e}},\quad \mathrm{M=D2,C2a,C2b},
\end{align} 
where $\tau_{\rm D2,C2a}=\tau_3$, $\tau_{\rm C2b}=\tau_4$, and $X_{\rm
 M}(t)$ 
 are smoothly varying functions, whose detailed features follow from the results of the Appendix~\ref{app:Det}. 
 Here we note that the overall signal strength $X_{\rm M}$ is by a factor $(E\tau/\hbar)^{1-d} \ll 1$ smaller
 than the strength function $X_{\rm C1}$ of the C1 process and in this smallness
 reflects the relatively smaller phase volume available to the returning of
 higher-order path topologies. 
  
 Summarizing the discussion so far, Fig.~\ref{PeakChronology} shows a typical chronology of echo signals in response 
 to two applied pulses as a sequence of dots of varying strength and
 angular orientation. The latter refers to directional information encoded in
 momentum space, to be discussed next.

\section{Momentum space echoes}
\label{MSE}

Although the essential classification 
of the system response in terms of echo times $\{\tau_i\}$ and 
corresponding path structures is universal, additional information can be
obtained if observables different from the coordinate projectors $\hat O=|\br
\rangle \langle \br|$ are chosen. Specifically, in this section we
turn to the complementary limit of momentum projectors, $\hat O = |\bk\rangle
\langle \bk|$, and look for echo signals in the scattering probability from $\bk$ to $\bk'$. Since now initial and 
final momentum are fixed, the formal loop order is decreased by one compared to the real-space setting. 
Namely, an $n$-mode contribution to the momentum-space signal will be made of $(n-1)$ momentum integrals, 
and the corresponding $n$-loop signal in real space is recovered by one supplementary momentum integration. 
Also, quantum coherence in momentum space no longer constrains the initial and final positions, but instead 
requires an alignment of initial and final momenta,
$\bk'=\pm\bk$. Therefore, in a momentum resolved scattering experiment, the $\mathrm{C1}$
echo is observed as a contribution to the backscattering
probability at $\bk'=-\bk$. In contrast, two-mode contributions will peak in the \emph{forward} direction $\bk=\bk'$. 
Forward-scattering coherence has been recently identified as particularly interesting in connection with
the onset of strong localization~\cite{Karpiuk2012,Micklitz2014,Lee2014}. 

\begin{figure}[b]
\centering
\includegraphics[width=0.45\textwidth]{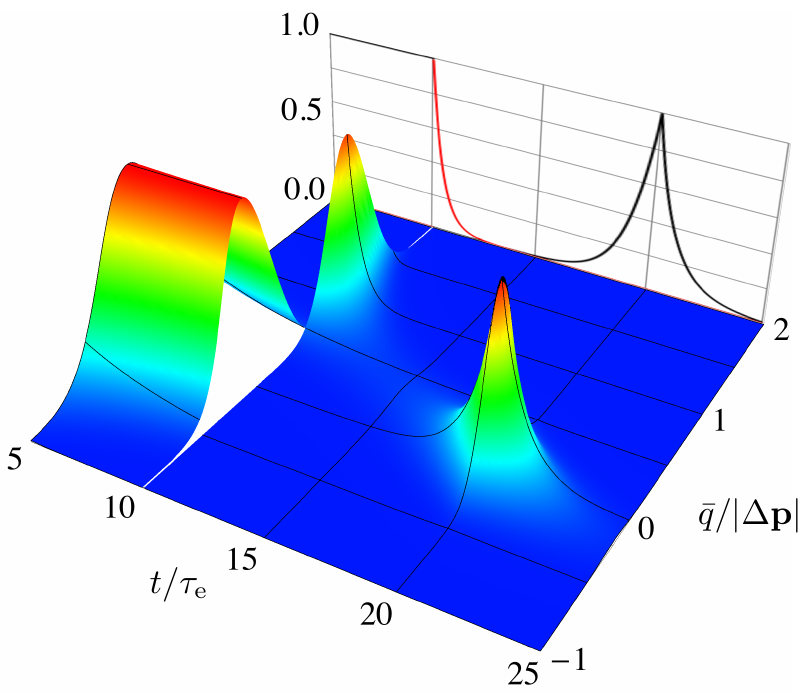}
\caption{\label{c1echoplot3d}
Single-mode Cooperon echo contrast $\delta X_\text{C1}(t,\bar\bq)/X_0$, given by \eqref{textonelooppeak}, 
as function of time $t$ and momentum $\bar{q}$ 
[in units of $\taue$ and $\Delta p$ and for $\bar{\bq}\parallel \delp$], with a dephasing pulse at $t_1=10\taue$. 
The momentum kick $\delp$ initially displaces the entire momentum distribution, then dephasing sets in, 
and the signal only revives at the echo time $\tau_1=2t_1$. 
The echo contrast at exact backsattering $\bar{q}=0$ is shown as the black curve in the side panel. 
The real-space signal \eqref{singleCooperonecho} follows after integration over $\bar\bq$. 
}
\end{figure}

In fact, coherence echoes in momentum space show a somewhat richer structure beyond the general forward and backward orientation. 
It is instructive to look first at the single-Cooperon coherent backscattering echo. Postponing a systematic derivation to the next 
Section~\ref{sec:FT}, we state here merely the expected result \cite{Cherroret2012}:   
\begin{align}
\label{app:cbse}
\delta X_\text{C1}(t,\bar\bq)
&=
X_0 \Pi_\text{C}(\bar{\bq},t,0,t), 
\end{align}
where 
$X_0= {4\pi \over  \nu}\tilde{\delta}(\epsilon_\bold{k}-\epsilon_{\bold{k}'})$ with 
$\tilde\delta(\epsilon)=\frac{\hbar}{2\pi \tau}\frac{1}{\epsilon^2+(\hbar/2\tau)^2}$ is a
broadened $\delta$-function keeping the arguments of the correlation function
on-shell, and $\Pi_\text{C}(\bar{\bq},t,0,t)$ is the Cooperon solution of the generalized diffusion equation 
\eqref{eq:DiffusionA} at momentum $\bar \bq = \bk+\bk'$ away from the backscattering direction.  

In the absence of external potentials, the simple diffusion equation is solved by the Gaussian  
$\delta X_\text{C1}(t,\bar\bq)
= X_0 e^{-Dt\bar{\bq}^2/\hbar^2}$, as recently predicted for a cold atom set up~\cite{Cherroret2012} 
and consequently observed~\cite{CBSexperiment}.  
Now, in presence of the  
dephasing \eqref{dephasingpot},  
the Cooperon takes the form   
\begin{align}
\delta X_\text{C1}(t,\bar\bq)
&= 
X_0 e^{-{D\over\hbar^2}\left(t\bar{\bq}^2 - 2 \bar\chi_1(t) \bar{\bq}\cdot\delp +  
  \bar\chi_2(t) \delp^2\right)}.
\label{textonelooppeak}
\end{align}
A derivation of this result, including expressions of the auxiliary functions $\bar{\chi}_{1/2}$ for general pulse profiles 
$f(t)$ (eqs.~\eqref{bchi1} and \eqref{bchi2}) can be found in Appendix~\ref{app:Diff} below.  
For a single $\delta$-pulse at time $t=t_1$, these functions simply vanish before the pulse and are 
 $\bar\chi_1(t)= (2t_1-t)$ and 
$\bar\chi_2(t)= | 2t_1-t |$ at times $t>t_1$ after the pulse. 
In that case, the echo contrast in the exact backward direction $\bar\bq=0$ reads 
\begin{equation} 
\delta X_\text{C1}(t,0)= X_0 e^{-\theta(t-t_1)|t-2t_1|/\taue}, 
\end{equation}
and thus describes a sharp drop at the pulse time $t_1$ followed by an exponential revival at the echo time 
$\tau_1=2t_1$. The corresponding real-space echo \eqref{singleCooperonecho} follows 
upon integration of eq.~\eqref{textonelooppeak} over $\bar\bq$.   

On the scale of $\delp$, the momentum-space signal shows a rather interesting dynamics,  as plotted in
Fig.~\ref{c1echoplot3d}. Initially, the dephasing kick displaces the entire momentum distribution by $\delp$, 
and thus also displaces the backscattering peak, which subsequently takes a finite time $\taue$ to dephase. 
As time increases, the point of highest contrast is found at $\bar\bq_0(t)
= \bar\chi_1(t)\delp/t$. It thus 
moves from $\bar\bq_0(t_1) = \delp$
towards the original position $\bar\bq_0(2t_1) =0$, 
reached at the echo time, and then continues onward to $\bar\bq_0 \to - \delp$ at long times. 
 The peak is severely suppressed at generic times, but
revives with perfect contrast at $t=2t_1$ where
$\bar\chi_1=0=\bar\chi_2$. 

For pulses of finite resolution in time, but still symmetric around
$t_1$, the peak always reaches the original
position $\bar\bq_0=0$ at $2t_1$ 
where the contrast penalty vanishes, implying a perfect revival, as a consequence of the 
general expressions \eqref{bchi1} and \eqref{bchi2}. Only for an asymmetric pulse the
contrast penalty generically remains finite, 
and the echo will appear with reduced contrast.  

In contrast to the C1 echo discussed so far, the higher-order processes D2 and C2 
show echoes in response to bi-temporal pulsing in the \emph{forward} scattering direction. 
A detailed discussion of the intricacies of momentum-resolved two-mode echoes can be found in Appendix~\ref{app:Det}. 
We first complete the general development of the theory by a systematic derivation of echo contributions via a field-theoretical approach.

\section{Field theory}
\label{sec:FT}

In this section we derive the results discussed so far  
within the framework of the diffusive nonlinear $\sigma$-model~\cite{efetov}. 
Compared to a direct perturbative
`diagrammatic' calculation, the $\sigma$-model greatly simplifies the handling
of the vertex regions distinguishing individual echo contributions. It also
`automatizes' the identification of echo time structures, which in a
diagrammatic framework have to be anticipated from the beginning. 
We use a simplified Keldysh version of the 
model~\cite{AndreevKamenev,LevchenkoKamenev,AltlandKamenev}, 
which is tailored to treat time dependent phenomena, and   
proceed to show how the theory yields the discussed echo structures. 
We invite readers not interested in technical details to skip this section and to proceed to 
``Summary and experimental realization".

\subsection{Effective theory}

Central to our discussion is a functional-integral partition function 
\begin{align}
	\label{Z1}
	\mathcal{Z}=\int DQ \,\exp(iS[Q]),
\end{align}
with effective action
\begin{align}
\label{s}
iS[Q]
&={\pi\nu\hbar\over 8}
 \int d\br\,
{\rm tr}\left(
2\partial^\phi Q(\br)
-
D(\partial^{\bold{a}}Q(\br))^2 
\right),
\end{align}
describing quantum diffusion on time scales, $t\gg\tau$, and eventually
Anderson localization on asymptotically large scales. 
Here, $\nu$ is the density of states per volume, $\tau$ the elastic scattering time, 
and 
$Q=\{Q^{\alpha\alpha'}_{ss',tt'}\}$ is a unitary matrix field,
$Q^{-1}=Q^\dagger$, bi-local in time $t,t'$ and two-dimensional in two
auxiliary spaces of Keldysh ($\K$) and time reversal ($\T$) variables, respectively. The $\K$-space indices
$\alpha=\pm$ discriminate between retarded $(+)$ and advanced $(-)$
propagators. The $\T$-space indices $s=\pm$  track time
reversal operations. For example, the matrix block $Q^{+-}_{+-}$ describes an
interfering pair of retarded and advanced amplitudes which are 
counter-propagating in time, $Q^{+-}_{++}$ describes interference of co-propagating
amplitudes, etc. The trace `tr' in \eqref{s} includes summation over all
indices, including continuous time, $\mathrm{tr}(A) =\sum_{\alpha,s}\int dt\,
A^{\alpha\alpha}_{ss,tt}$. Likewise, matrix multiplication is defined as
$(AB)^{\alpha \alpha'}_{ss',tt'} =\int dt^{\prime\prime}
\sum_{\alpha^{\prime\prime},s^{\prime\prime}} A^{\alpha
\alpha^{\prime\prime}}_{ss^{\prime\prime},tt^{\prime\prime}} B^{\alpha^{\prime
\prime}\alpha^\prime}_{s^{\prime\prime}s',t^{\prime\prime}t'}$. With these
conventions, the matrix field $Q$ is defined to obey the nonlinear constraint
\begin{align*}
	(Q^2(\br))_{tt'}=\openone\delta(t-t'),
\end{align*}
where $\openone$ is the unit-matrix in $\K\otimes\T$. 
Invariance under time-reversal reflects in a second constraint
 \begin{align}
\label{scq}
(Q_{tt'})^\trsp =  \sigma_2^{\T} Q_{-t',-t} \sigma_2^{\T},
\end{align} 
where $\trsp$ is transposition in $\K\otimes\T$ and the Pauli matrices 
$\sigma_i^{\mathrm{X}}$ act in $\mathrm{X}=\K,\T$-space, respectively.

The particle matrix field $Q$ couples in Eq.~\eqref{s} to the external fields via the covariant
derivatives 
\begin{align}
	\label{CovDer}
	\partial^\phi Q(\br)_{ss',tt'} 
	& = 
	[\partial_t-\partial_{t'} + i \phi(\br, s t)+i\phi(\br,s't')]Q_{ss',tt'}(\br),\crcr
	\partial^\bold{a} Q(\br)_{ss',tt'} 
	& =  
	[\partial_\br+i s\bold{a}(\br,st)-i s'\bold{a}(\br,s't')]Q_{ss',tt'}(\br).
\end{align}
The covariant form of these derivatives reflects the local U(1) gauge invariance of the theory. 
The sign structure in $\T$-space ensures that the covariant derivatives of the $Q$-field are 
consistent with the time reversal condition~\eqref{scq}.

The action \eqref{s} is manifestly invariant under `rotations' $Q_0 =
\sigma_3^\K \mapsto T_0
\sigma_3^\K T_0^{-1}$, where $T_0=\mathrm{const.}$ is a matrix in
$\K$-space  and $Q_0=\sigma_3^{\mathrm{K}}$ a saddle point not containing
interference terms, $(\sigma_3^\K)^{+-}= (\sigma_3^\K)^{-+}=0$. This saddle point
describes the system before the appearance of diffusion modes, the sign
structure in $\mathrm{K}$-space being a consequence of Green function
causality.~\cite{AndreevKamenev,LevchenkoKamenev,AltlandKamenev} 
Transformations with $T_{tt'}(\br)$ slowly varying in space and
time generates soft `Goldstone modes' that 
represent physical diffusion modes, much like small $(\br,t)$-dependent
-rotations of the spins in a ferromagnet describe magnon modes. We
therefore parametrize the relevant nonlinear field manifold by 
\begin{align}
\label{spp}
Q(\br)&= 
 T(\br) \sigma_3^\K T^{-1}(\br) 
\end{align}
and with smooth fluctuations $T$.

\subsection{Cooperon and diffuson modes}

To explore the effect of soft mode fluctuations, we parameterize the rotation
matrices as $T=e^{W/2}$ where the generators $W$ are chosen to
anti-commute with the saddle point, $\big[ \sigma_3^\K, W \big]_+ =
0$. These generators are block off-diagonal in $\K$-space,
\begin{align}
	\label{Wstructure}
	W=\left(\begin{matrix}
		&B\cr -B^\dagger&
	\end{matrix}\right)_\K,\qquad B=\{B^{ss'}_{tt'}\},
\end{align}
their anti-hermitean structure required by the unitarity of $Q$. 
The time reversal symmetry relation \eqref{scq} implies 
$(W_{tt'})^\trsp =  -\sigma_2^{\T} W_{-t',-t} \sigma_2^{\T}$. For the $B$-matrices this means
\begin{align}
 \label{symrel}
	B^{--}_{tt'}=\bar{B}^{++}_{-t',-t},\qquad
	B^{-+}_{tt'}=-\bar{B}^{+-}_{-t',-t},
\end{align}
where the overbar is complex conjugation. We will identify modes $B^{\pm\pm}$ 
of identical ($B^{\pm\mp}$  of opposite)
time orientation of amplitudes as diffuson (Cooperon) modes, and define 
$	B^{++}_{tt'}\equiv D_{tt'}$ and $B^{-+}_{tt'}\equiv C_{tt'}$ 
or in $\T$-space 
\begin{align}
	\label{Bexplicit}
		B_{tt'}=\left(\begin{matrix}
		D_{tt'}& -\bar C_{-t'-t}\crcr
		C_{tt'} &\bar{D}_{-t'-t}
	\end{matrix}\right).
\end{align}

The strategy now is to substitute the expansion 
\begin{align} \label{Qexpansion}
Q
&=T \sigma_3^\K T^{-1}
\simeq 
\sigma_3^\K \left( 1- W +  W^2/2 +\dots\right)
\end{align}
into the action \eqref{s} and to expand in $W$. There is no
zeroth-order contribution, and the first order vanishes around the
saddle point.  To second order, the action decouples into two
quadratic actions for diffuson and Cooperon, respectively, 
\begin{widetext}
\begin{align}
iS^{(2)}_\text{D} 
&= 
-\frac{\pi\nu\hbar}{2} 
\int dt \int dt' \int d\br \, 
D_{t't}(\br)
\left( \partial_t + \partial_{t'} 
+ 
i\left[ 
\phi(\br,t) - \phi(\br,t') 
\right] 
-
D \left(
\partial_\br
+i \left[
\bold{a}(\br,t) - \bold{a}(\br,t') 
\right]
\right)^2 
\right) 
\bar{D}_{tt'}(\br), \\
iS^{(2)}_\text{C} 
&= 
-\frac{\pi\nu\hbar}{2} 
\int dt \int dt' \int d\br \,
C_{-t't}(\br)
\left( \partial_t - \partial_{t'} 
+ 
i\left[ 
\phi(\br,t) - \phi(\br,t') 
\right] 
-
D \left( 
\partial_\br
+i \left[
\bold{a}(\br,t) + \bold{a}(\br,t') 
\right]
\right)^2 
 \right) 
\bar C_{t,-t'}(\br).
\end{align}  
\end{widetext}
The kernels are just the differential
operators~\eqref{eq:DiffusionA}. The  
correspondence with the diffusion modes $\Pi_\text{D/C}$ can be made
more explicit by calculating the expectation values $\left\langle \dots \right\rangle^{(2)} \equiv \int
D(D,C)\exp[iS^{(2)}_\text{C}+iS^{(2)}_\text{C}](\dots)$ 
with the quadratic action, where
$D(D,C)$ stands for integration over the matrix-fields
$D,C$. Since a complex Gaussian integral yields the 
inverse of the action kernel, the expectation values 
\begin{widetext}
\begin{align}
\label{cdprop}
\langle
\bar{D}_{t^+,t^-}(\br)D_{t^- -T, t^+ -T'}(\br')
\rangle^{(2)}
&=
{ 2\over \pi\nu\hbar }
\Pi_\text{D}(\br,\br';t^+,t^-,T)\delta(T-T'),
\nonumber
\\
\langle
\bar C_{t^+,-t^-}(\br)C_{-t^- -T, t^+ -T'}(\br')
\rangle^{(2)}
&=
{2\over \pi\nu\hbar}
\Pi_\text{C}(\br,\br';t^+,t^-,T) \delta(T-T'),
\end{align}
\end{widetext}
obey Eqs.~\eqref{eq:DiffusionA} and thus are
identical to the modes considered there.  
For later reference, we note that the $\delta(T-T')$-functions above  
are regularized to the shortest time scales $\sim \tau$ resolved by the field theory; 
they are to be understood as broadened Lorentzians with finite peak height $\delta(0)=\tau^{-1}$.
(For completeness, we note that the Gaussian integrals are unit normalized,
$\left\langle 1\right\rangle =1$, i.e. they do not yield a non-trivial
`functional determinant'. The physical principle behind this is Green function
causality, which implies the unit-valuedness of the determinants of the
Cooperon and diffuson differential operators. For further discussion of this
point, we refer to Refs.~[\onlinecite{AltlandKamenev,LevchenkoKamenev}].)

\subsection{Generation of observables}

Starting from this section, we focus on 
momentum-space coherences. 
As already discussed in Sec.~\ref{MSE}, momentum-resolved correlations provide additional information about 
the parity of interference processes under time-reversal, which 
complements the information contained in spatial correlations. 
A generalization of the formalism to the generic coherent states introduced in 
section~\ref{sec:semiclassical_approach_to_coherence_echoes} is straightforward.
The relevant correlation
function~\eqref{eq:Xdef} then is 
\begin{equation}
\label{Xkk}
X(t,\bold{k},\bold{k}') = \left\langle \left|\langle{\bk'}|\exp \{
    -iHt \}|\bk\rangle  \right|^2\right\rangle, 
\end{equation}
namely the ensemble-averaged scattering probability from $\bk$ to $\bk'$ in time t. 
In order to compute this correlation function from the field theory, we introduce two source parameters 
$\alpha= \{\alpha^\text{i},\alpha^\text{f}\}$ together with the projectors 
$\mathcal{P}^\text{i/f}_{ss',tt'}(\bp)=
	\delta(\bk_\text{i/f}-s \bp)\delta(t_\text{i/f}-s t)\delta_{ss'}\sigma^\K_\pm$ in time and momentum, 
where the external time and momentum 
	arguments are 
\begin{align}
t_\mathrm{i} &= 0, &  t_\mathrm{f}&=t,	\label{tIdentification}\\
\bk_\mathrm{i} & =\bk, &  \bk_\mathrm{f}&=\bk',   \label{kIdentification}		
\end{align} 
and $\sigma_\pm^\K = \frac{1}{2}(\sigma^\K_1 \pm i \sigma^\K_2)$ are raising and lowering operators in $\K$-space. 
The source-augmented action $S[Q,\alpha]=S[Q]+S_\alpha[Q]$ is given by 
$S[Q]$ of \eqref{s} and the sum $S_\alpha[Q] = S^{\rm I}_\alpha[Q] + S^{\rm II}_\alpha[Q]$
of two contributions, one linear and the other quadratic in the sources,
\begin{align}
\label{Salphalin}
S_{\alpha}^{\rm I}[Q]
&=   
 {1\over 2} 
 \int d\br   \,
 {\rm tr}
 \left[
 Q(\br)\left(\alpha^\mathrm{i}\mathcal{P}^\text{i}+\alpha^\mathrm{f}\mathcal{P}^\mathrm{f}\right)
 \right], \\
S_{\alpha}^{\rm II}[Q]
&=
{\alpha^\mathrm{i}\alpha^\mathrm{f} \over 2i} 
\int (dp)\int (dp') {\rm tr}
\left[ 
 Q_{\bold{p}-\bold{p}'}\mathcal{P}^\text{i}(\bp')
 Q_{\bold{p}'-\bold{p}}\mathcal{P}^\text{f}(\bp)
\right].
\label{Salphaquad}
\end{align}
Here, $\mathcal{P}_{ss',tt'}^\text{i/f}=
	\delta(t_\text{i/f} -s t)\delta_{ss'}\sigma^\K_\pm$ without momentum argument projects only in 
time. Further, 
$Q_{\bp}$ is the Fourier transform of $Q(\br)$. 
The correlation function \eqref{Xkk} is then obtained by twofold differentiation
of the generating partition functional $\mathcal{Z}[\alpha]=\int
DQ\,\exp(iS[Q,\alpha])$, 
\begin{align}
\label{fid}
X(t,\bold{k},\bold{k}') 
&=-2\pi^2\hbar \tilde
\delta(\epsilon_\bk - \epsilon_{\bk'})
\partial_{\alpha^\mathrm{f}}\partial_{\alpha^\mathrm{i}}
\left.{\cal Z}[\alpha]\right|_{\alpha=0}.  
\end{align} 
Here,  
$\tilde\delta(\epsilon)=\frac{\hbar}{2\pi \tau}\frac{1}{\epsilon^2+(\hbar/2\tau)^2}$ is a
broadened $\delta$-function keeping the arguments of the correlation function
on-shell.

\subsection{Echo spectroscopy in momentum space}
\label{AppLES} 

Based on a systematic expansion in diffusion modes, we can now express the echo signals 
in a fully quantitative manner.  
We here concentrate on momentum-resolved correlation functions
and recall that corresponding signals in real space are generated by integration over the 
remaining momentum argument.
The strategy is to substitute the expansion~\eqref{Qexpansion} into the
source terms, to differentiate w.r.t. external parameters
$\alpha^\mathrm{i,f}$ and to compute the ensuing Gaussian integrals with the
help of~\eqref{cdprop}.  To the individual contributions obtained in this way, we
may attribute a topology and in this way establish contact to the
semiclassical representations of
section~\ref{sec:semiclassical_approach_to_coherence_echoes}.

\subsubsection{Classical relaxation}
\label{sec:zeroloopecho} 

To lowest order, the field theory
reproduces the classical, ergodic 
spread of the population over 
the entire energy shell. 
This is found by expanding the source \eqref{Salphalin} to linear
order in $W$. Substituting the expansion~\eqref{Qexpansion} and using 
Eqs.~\eqref{Wstructure} to~\eqref{Bexplicit} to represent the internal structure of the $W$-generators, 
a straightforward computation shows
\begin{align}
\label{W1}
S_\alpha^{\rm I,1}
 &=  
- 
\left[
\alpha^\mathrm{i}  D_{t_\mathrm{i}t_\mathrm{i}}(0) 
+ 
\alpha^\mathrm{f} \bar{D}_{t_\mathrm{f}t_\mathrm{f}}(0)
 \right], 
 \end{align}
where the arguments in parentheses 
refers to zero momentum $\bold{q}=0$. Fixing time arguments,
Eq.~\eqref{tIdentification}, and differentiating w.r.t.\ sources, we obtain the contribution 
 \begin{align}
\label{f0a}
X_0(t,\bk,\bk')
&=
2\pi^2\hbar \tilde{\delta}(\epsilon_\bold{k}-\epsilon_{\bold{k}'})
\langle
 \bar{D}_{tt}(0)
D_{00}(0) 
\rangle
\nonumber \\ &
=
{4\pi \over  \nu}\tilde{\delta}(\epsilon_\bold{k}-\epsilon_{\bold{k}'})  \Pi_{\rm D}(0;t,t,t)
 \end{align}
 to the correlation function $\eqref{Xkk}$, where the first argument of $\Pi_\text{D}$ refers to $\bq=0$ momentum.  
Eq.~\eqref{f0a} is
structureless on the momentum shell $|\bold{k}|\approx|\bold{k}'|$ and thus 
describes the terminal state of classical momentum shell relaxation, reached at time scales larger than the scattering time. 
(The dynamics on shorter time scales $t\sim\tau$ can be resolved
by a master equation \cite{Plisson2012}.{}) 
Since the simple diffuson
$\Pi_\text{D}(0,t,t,t) = \theta(t) e^{-Dt\bold{q}^2/\hbar^2}|_{\bq=0}=\theta(t)$ 
is insensitive to dephasing, the isotropic
background \eqref{f0a} is 
\begin{align}
\label{f0}
X_0&= 
{4\pi \over  \nu}\tilde{\delta}(\epsilon_\bold{k}-\epsilon_{\bold{k}'}) 
 \end{align}
at all times $t\gg \tau$, and this independently of external dephasing
potentials $\phi(\br,t)$.

\subsubsection{Single-mode backscattering echo}
\label{sec:oneloopecho} 

The leading order coherence signal is the backscattering 
peak of Refs.~\onlinecite{Cherroret2012,CBSexperiment}. 
This term is generated by inserting the quadratic contribution in generators into 
the quadratic source Eq.~\eqref{Salphaquad}. There are two qualitatively different 
types of terms, arising from the expansion of (i) both $Q$ matrices to 
linear order in $W$ and (ii) one $Q$-matrix to second and the other to zeroth order in $W$. 
However, only type (i) gives a finite contribution.  
Performing the twofold derivative \eqref{fid} and inserting the explicit parametrization 
one arrives at contributions from diffuson and Cooperon modes. 
Only the latter give a finite expectation value
\begin{align}
\label{W2c1}
\delta X_\text{C1}(t,\bar\bq)
&=
2\pi^2\hbar \tilde{\delta}(\epsilon_\bold{k}-\epsilon_{\bold{k}'})
\langle 
\bar{C}_{t 0}(\bar\bq) C_{-t0}(-\bar\bq)
\rangle, 
\end{align}
where $\bar\bq=\bold{k}'+\bold{k}$ denotes 
 the deviation from exact backscattering. 
 Upon inserting the propagator \eqref{cdprop} 
  one finds the contribution eq.~\eqref{app:cbse} of section~\ref{MSE}.

\subsubsection{Double-mode forward scattering echo}
\label{sec:twoloopecho} 

The lowest order contribution to the forward scattering peak appears in quartic order in generators $W$
in the quadratic source term. Again there are various contributions and we only give here the relevant term, resulting 
in non-vanishing contribution to the observable of interest.  
 Following the same steps as in the single-mode contribution, i.e.
performing the two fold derivative and 
inserting the explicit parametrization of generators 
one arrives at the following two contributions from diffuson and Cooperon modes 
(for simplicity we suppress the momentum arguments for the moment and only state those contributions
with a finite expectation value),
\begin{widetext}
\begin{align}
\label{w4modes}
\delta X_2(t)
&=
2\pi^2\hbar \tilde{\delta}(\epsilon_\bold{k}-\epsilon_{\bold{k}'})
 \int dt' \int dt'' \, 
\langle
\bar{D}_{tt'} D_{t'0} 
D_{0t''} \bar{D}_{t''t} 
+ \bar{C}_{t-t'} C_{-t'0} 
\bar{C}_{t''0} C_{-tt''} 
 \rangle.
\end{align}
\end{widetext}
Reintroducing momenta dependencies and 
inserting the propagators \eqref{cdprop}
we arrive at the two-mode contributions 
\begin{align}
\label{delX2}
\delta X_2(t,\bq)
&=
\delta X_\text{D2}(t,\bq)
+\delta X_\text{C2}(t,\bq)
,
\end{align}
where $\bq= \bk' - \bk$ denotes the deviation from forward scattering.  
The two-diffuson contribution reads
\begin{align}
\label{D2oft}
\delta X_\text{D2}(t,\bq)
&
=
{2X_0\over \pi\nu \hbar}
\int dt' \int (dq') \,
\Pi_\text{D}(\bold{q}'+\bq,t-t',t,t-t')
\nonumber \\
&\quad \quad \quad 
\quad \quad \quad \times
\Pi_\text{D}(\bold{q}',t,t',t'),
\end{align}
with $(dq)=d\bold{q}/\hbar^d$, 
and 
\begin{align}
\label{C2oft}
\delta X_\text{C2}(t,\bq)
&=
{2X_0\over \pi\nu \hbar}
\int dt' \int (dq') \,
\Pi_\text{C}(\bold{q}'+\bq,t, t',t-t')
\nonumber \\
&\quad \quad \quad 
\quad \quad \quad \times
\Pi_\text{C}(\bold{q}',t',0,t')
\end{align}
is the two-Cooperon contribution. The resulting 
coherent forward scattering echo in momentum space is discussed in detail in Appendix~\ref{app:Det}. 
By a momentum-integration over $\bq$, one arrives at the two-loop echoes 
 in real space discussed in Section~\ref{HigherOrder}.

\subsubsection{Higher order diffusion modes}

In the absence of dephasing pulses and in $d\leq 2$ dimensions, the proliferation of quantum diffusion modes 
eventually results in strong, Anderson localization. Using non-perturbative methods,  
the resulting temporal builtup of the forward scattering peak 
has been recently calculated in a
quasi-onedimensional geometry and with a weak magnetic field breaking
time-reversal symmetry~\cite{Micklitz2014}. 
In principle, it is possible to also push echo spectroscopy to 
higher order, extending the theoretical analysis
above to $n>2$-pulse dephasing trains. Indeed, for $n>2$ pulses, convolutions of diffuson and
Cooperon modes begin to appear, and the detection of those would provide a
highly non-trivial test of our present understanding of the dynamical
processes that result in Anderson localization. 
The systematic investigation of echo times $\tau_i$ resulting from $k$-mode contributions 
is an interesting, though at the present stage theoretical problem 
that we leave for future investigations.

\section{Summary and experimental realization}
\label{Conclusion}

In summary, the proposed echo spectroscopy 
provides a highly resolved probe into the interference processes fundamental
to quantum localization. Such type of diagnostics is essential in situations
where it is difficult to separate coherent from classical
backscattering~\cite{ClassicalCBS1,ClassicalCBS2}, or to distinguish between strong Anderson
localization and classical potential trapping. Unlike indiscriminate
dephasing, echo spectroscopy permits to distinguish whether or not certain
coherent processes rely on anti-unitary symmetries such as time reversal
invariance. While the detection of echoes becomes increasingly demanding with
the number of diffusive modes involved, measuring the peak heights and widths
of the discussed lowest order signals would quantitatively determine the phase
space volume available to fundamental coherent scattering processes.

For a concrete realization, we suggest to use the 
`disorder quench' protocol with ultracold gases~\cite{CBSexperiment}. In
this variant, a Bose-Einstein
condensate is released from a trap and let to evolve in a far-detuned optical speckle
field for some time, after which real-space \cite{Billy2008} or momentum
\cite{CBSexperiment}  distributions are measured. The advantage of this
setup is that it (i) allows to prepare well-defined initial wave packets with
small spread around finite $\bp$ and (ii) that the atoms are suspended against
gravity by a magnetic field gradient which can be changed below
the ms time-scale of $\tau$ to impart the dephasing kicks. A concrete
realization, therefore, seems immediately possible within at least one existing setup.
And indeed, at the single-pulse
level, first experimental results are already available
\cite{EchoExp}.  
 The observation of quantum interference processes higher than first 
order within echo spectroscopy may be experimentally challenging but is arguably
realistic using similar setups, possibly constrained to lower-dimensional
geometries where return probabilities are enhanced and echo amplitudes thus 
larger. It is straightforward to push the theoretical analysis to
$n$-pulse trains and for $n>2$, processes relying on convolutions of diffuson and
Cooperon modes begin to appear. We are not aware of experiments systematically 
probing the onset of Anderson localization beyond single-Cooperon backscattering. 
The detection of higher mode echoes would, therefore, provide a highly non-trivial test 
of the validity of our conceptual understanding of Anderson localization. 
Experimental resolvability being the key limiting factor, it seems reasonable to stay at the $n=2$ 
level for the moment. 

Another interesting avenue would be the `in silico' echo spectroscopy
of many body localization processes~\cite{ManyBodyCBS,MBSpinecho,MBLinterference}. 
At this point, even very basic aspects of
the phenomenon -- such as the effective dimensionality of the underlying
stochastic dynamics, the principal applicability of diffusion mode approaches
in Fock space, etc. -- are not very well understood, and the detection of
echoes in response to external pulses might provide valuable insights.

{\it Acknowledgements:---}T.~M. gratefully acknowledges useful discussions
with H.~Micklitz and support by Brazilian agencies CNPq and FAPERJ. 
C.A.M. acknowledges hospitality of Universit\'e 
Pierre et Marie Curie and Laboratoire Kastler Brossel, 
Paris. A.A. acknowledges support by SFB/TR 12 of the Deutsche
Forschungsgemeinschaft

\begin{appendix}

\section{Coherent forward scattering echo}
\label{app:Det}

In a perturbative mode-expansion the leading contribution 
to the forward scattering peak in the momentum correlation function
results from the two-mode contributions Eq.~\eqref{delX2}. 

Without dephasing, diffuson and Cooperon are equal, 
$\Pi_\text{C/D}(\bold{q},t^+,t^-,T)
=  e^{-D\bold{q}^2 T/\hbar^2}\theta(T)$,  
and the forward scattering peak is readily found upon
Gaussian integration over the intermediate momenta $\bq'$ in \eqref{D2oft} and \eqref{C2oft}, 
\begin{align}
\delta X_2(t,\bq)
={4 X_0 \pi^{d/2} \over \pi \nu \hbar (D t)^{d/2}}
\int_0^t dt'  
e^{-{Dt'\over\hbar^2}(1-t'/t)\bq^2}. 
\end{align}
In the forward direction $\bq=0$, this
yields 
\begin{align}
\label{ftdx2}
\delta X_2(t)
&=
{4 \pi^{d/2} X_0 t\over \pi \nu \hbar (D t)^{d/2}}
\sim
t^{(2-d)/2}, 
\end{align}
which in the $d=2$ weak localization regime is a constant contribution of order $1/kl$. \cite{Karpiuk2012}  
With a single pulse, the two-mode terms provide a smooth background without particular structure 
in time or momentum. We therefore turn directly to the effect of two dephasing pulses, applied at times $t=t_1$ and
$t=t_2>2t_1$, which select the resonant signal characteristic of the two-mode contributions.

\subsection{Two-mode diffuson D2}
\label{app:D2echo}

First we study the two-mode diffuson $\delta X_\text{D2}(t)$, eq.~\eqref{D2oft}. 
Inserting the general solution \eqref{appPiD} and integrating over $\bq'$ yields 
\begin{align}
\label{DelXD2int}
\delta X_\text{D2}(t,\bq)
&=
{2\pi^{d/2} X_0\over \pi\nu \hbar(Dt)^{d/2}}
\int_0^t dt' \,
e^{-\phi_\text{D2}(t',t,\bq)},
\end{align}
with the contrast penalty
\begin{widetext}
\begin{align}
\label{phiD2def}
\phi_\text{D2}(t',t,\bq)
&= {D\over\hbar^2}\Big[
t' \left(1-\frac{t'}{t} \right) \bq^2 
+ 2\left\{
\left(1-\frac{t'}{t} \right)
\chi_1(t,t',t')-\frac{t'}{t}\chi_1(t-t',t,t-t')
\right\}
\bq\cdot\delp 
\nonumber\\
& \qquad
+ \left\{ 
\chi_2(t-t',t,t-t')
+
\chi_2(t,t',t') 
-
\tfrac{1}{t}
[\chi_1(t-t',t,t-t')
+
\chi_1(t,t',t') ]^2 
\right\}\delp^2 \Big].
\end{align}
\end{widetext} 
The echo signal properly speaking stems from the temporal configuration shown
in the left panel of Fig.~\ref{2ndOrderEchoReal}, where each diffuson mode contains 
exactly one pulse, and which is selected 
by choosing in \eqref{DelXD2int} the  
integration limits 
\begin{equation}
\label{D2echointerval}
\max(t_1,t-t_2)<t'<\min(t_2,t-t_1). 
\end{equation}  
Then, the pulse functions  
 become  (see Appendix~\ref{app:Diffuson} for details)
\begin{align}
\label{appchi1d}
\chi_1(t-t',t,t-t')
&=t_1-t_2+t',
\\
\chi_1(t,t',t') 
&=t_2-t_1+t'-t,
\end{align} 
as well as $\chi_2(t-t',t,t-t')=|\chi_1(t-t',t,t-t')|$ and $\chi_2(t,t',t')=|\chi_1(t,t',t')|$. 
Exactly in the forward direction $\bq=0$,  eq.~\eqref{phiD2def} reduces to 
\begin{align}
\label{Gttprimeq0}
\phi_\text{D2}(t',t,0)
&=
 \left\{
\chi_2(t-t',t,t-t')
+
\chi_2(t,t',t') 
\right\}
/\taue,
\end{align}
where we have also dropped the last term inside the parentheses in the
second line of \eqref{phiD2def} multiplying $\delp^2$, which is small for 
$\taue =\hbar^2/(D\delp^2)\ll t$.

\begin{figure}
\centering
\includegraphics[width=0.45\textwidth]{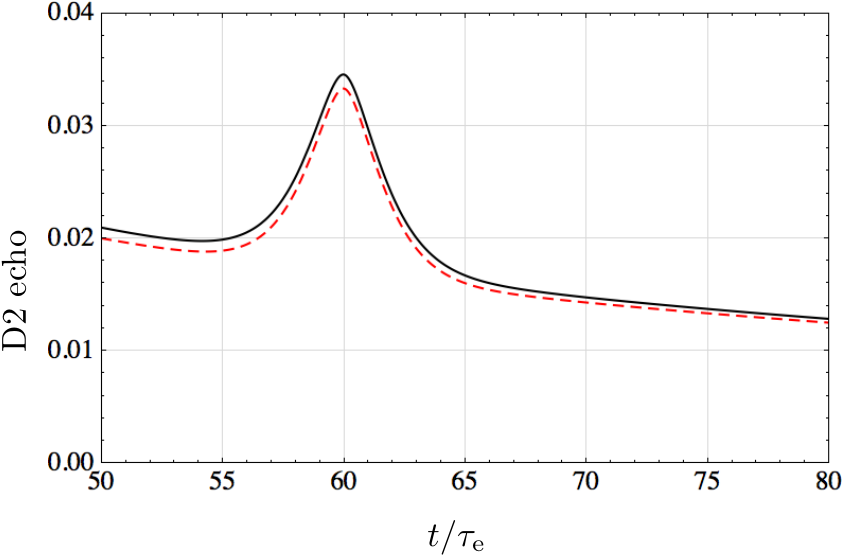}
\caption{\label{d2echoplotq0}
Two-mode diffuson signal $\delta X_\text{D2}(t,0)$ in the forward
direction relative to the non-pulsed contribution, i.e., half of
\eqref{ftdx2}, 
as function of $t/\taue$, after two dephasing pulses at $t_1 = 10 \taue$ and $t_2 = 40 \taue$. 
The black line shows the result of \eqref{DelXD2int}, a pronounced
echo around $\tau_3=60\taue$.  The red dashed curve shows the
analytical approximation, eq.~\eqref{echo3bis}, above the background.}  
\end{figure}

As function of $t$ around $\tau_3=2(t_2-t_1)$, the signal then is very
well approximated by 
\begin{align}\label{echo3bis}
\delta X_\text{D2}(t,0)
&=
{2\pi^{d/2} X_0\over \pi\nu\hbar (Dt)^{d/2}} 
\left( \taue + |t-\tau_3| \right) e^{-|t-\tau_3|/\taue}. 
 \end{align} 
 This echo signal is exponentially suppressed outside the echo time $\tau_3=2(t_2-t_1)$, showing a 
quadratic departure for $|t-\tau_3|\ll\taue$. 
At the echo time $t=\tau_3$, the signal remains smaller by a factor $\taue/t\ll1$ compared 
to the non-dephased signal \eqref{ftdx2}. This factor results from phase space reduction: 
without a field pulse the two
diffusons can connect at any time $0<t'<t$, while in presence of the 
two pulses the time $t'$ is effectively restricted to an interval of size
$\taue$ around $t'=2t_1$.

Figure~\ref{d2echoplotq0} shows the D2 contrast after two pulses at $t_1=10\taue$ and 
$t_2=40\tau2$ with its echo at $\tau_3= 60\taue$, relative to the non-pulsed diffuson signal, i.e., half of \eqref{ftdx2}. 
Actually, the echo contrast \eqref{echo3bis} appears on top of a smooth background, 
created by a combination of a double-pulsed diffuson with a non-pulsed diffuson. 
These contributions stem from the $t'$-integration outside the interval \eqref{D2echointerval} 
and result in a flat background of the same order than the echo itself. The black line shows the 
result of the full integration \eqref{DelXD2int}, whereas the dashed red line shows the analytical 
approximation \eqref{echo3bis} plus the background of unity.

\subsection{Two-mode Cooperon C2}

Next we turn to the two-mode Cooperon $\delta X_\text{C2}(t)$, eq.~\eqref{C2oft}. 
Using the general Cooperon solution \eqref{PiCqt} and integrating over $\bq'$ yields 
\begin{align}\label{DelXC2int}
\delta X_\text{C2}(t,\bq)
&=
{2\pi^{d/2} X_0\over \pi\nu\hbar (Dt)^{d/2}}
\int_0^t dt' \,
e^{-\phi_\text{C2}(t',t,\bq)},
\end{align}
where the contrast penalty now reads
\begin{widetext}
\begin{align}
\label{phiC2def}
\phi_\text{C2}(t',t,\bq)
&= {D\over\hbar^2}\Big[
t' \left(1-\frac{t'}{t} \right) \bq^2 
+ 2\left\{
\left(1-\frac{t'}{t} \right)
\bar\chi_1(t',0,t')-\frac{t'}{t}\bar\chi_1(t,t',t-t')
\right\}
\bq\cdot\delp 
\nonumber\\
& \qquad
+ \left\{ 
\bar\chi_2(t',0,t')
+
\bar\chi_2(t,t',t-t') 
-
\tfrac{1}{t}
[\bar\chi_1(t',0,t')
+
\bar\chi_1(t,t',t-t') ]^2 
\right\}\delp^2 \Big].
\end{align}
\end{widetext} 
The principal echo signal stems from the upper configuration in the
right panel of Fig.~\ref{2ndOrderEchoReal}, where each Cooperon mode contains 
exactly one pulse and which is selected 
by choosing in \eqref{DelXC2int} the  
integration limits 
\begin{equation}
\label{C2echointerval}
t_1 <t'< t_2. 
\end{equation}  
Then, the pulse functions  \eqref{barchi1} and \eqref{barchi2} become  
\begin{align}
\bar\chi_1(t',0,t')
&=2t_1-t',
\\
\bar\chi_1(t,t',t-t') 
&=2t_2-t-t',
\end{align} 
as well as $\bar\chi_2(t',0,t')=|\bar\chi_1(t',0,t')|$ and $\bar\chi_2(t,t',t-t')=|\bar\chi_1(t,t',t-t')|$.

\begin{figure}
\centering
\includegraphics[width=0.45\textwidth]{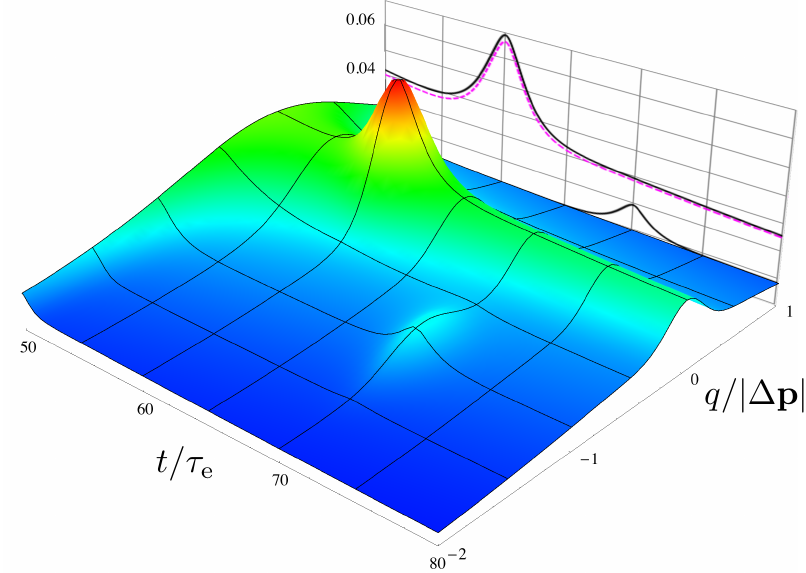}
\caption{\label{d2c2echoplot3d}
Two-mode echo contrast, eq.~\eqref{delX2}, relative to the non-pulsed signal  \eqref{ftdx2} as function 
of $\bq\parallel\delp$ and $t$ [in units of $\Delta p$ and $\taue$] after two dephasing pulses at $t_1 = 10 \taue$ 
and $t_2 = 40 \taue$. The principal echo of processes D2 and C2a appears in the exact forward direction at time 
$\tau_3=2(t_2-t_1)$, whereas the side echo C2b, Eq.~\eqref{C2bechoprofile}, appears shifted by $\delp$ and at 
later time $\tau_4=2t_2-t_1$.}  
\end{figure}

Exactly in the forward direction $\bq=0$,  eq.~\eqref{phiC2def} reduces to 
\begin{align}
\label{Cttprimeq0}
\phi_\text{C2}(t',t,0)
&=
 \left\{
\bar\chi_2(t',0,t')
+
\bar\chi_2(t,t',t-t') 
\right\}
/\taue,
\end{align}
where we have also dropped the last term inside the parentheses multiplying $\delp^2$, which is small for 
$\taue =\hbar^2/(D\delp^2)\ll t$.  
As function of $t$ around $\tau_3=2(t_2-t_1)$, the C2 echo signal is then identical to the D2 signal, eq.~\eqref{echo3bis}. 

This configuration is, however, not the only possible situation
where an echo can arise. Another possibility is shown in the bottom
part of the right panel in 
Fig.~\ref{2ndOrderEchoReal}. This produces an echo at finite momentum
$\bq=-\delp$ shifted slightly from the exact forward direction. The
reason is that each Cooperon is peaked at backscattering relative to
its incident and final momenta. But when the dephasing pulse hits the
first Cooperon near the end, this produces an enhancement at
intermediate momentum $\bk+\bk'' = \delp$ (this is already
seen in Fig.~\ref{c1echoplot3d} from the displaced C1 peak just
after the first kick). The second Cooperon is hit by the pulse in the
center and thus produces the usual backscattering enhancement at   
$\bk'+\bk''=0$. Altogether we expect a peak at 
$\bk'= \bk - \delp$ or indeed  $\bq=- \delp$. Its height is slightly smaller than the principal peak. 
To a very good approximation, the temporal peak profile is given by 
\begin{align} 
\label{C2bechoprofile}
  \delta X_\text{C2}(t,-\delp) & = \frac{2\pi^{d/2} X_0 \tau_\mathrm{e}}
  {\pi\nu\hbar (D t)^{d/2}} h\left(\frac{t-\tau_4}{\tau_\mathrm{e}}\right)
  \end{align}  with 
$h(s) = \frac{2}{3}\left[2e^{-|s|} - e^{-2|s|}\right]$. 
Remark that no such configuration is possible  for the D2 topology because there the loops are traversed in opposite order, 
and consequently it is impossible for the pulses to hit only one diffuson mode, but not the other.

Summarizing the double-mode momentum-space discussion, Fig.~\ref{d2c2echoplot3d} shows the combined signal, 
eq.~\eqref{delX2},  normalized with respect to the unperturbed signal \eqref{ftdx2}.

\section{Diffusion modes with dephasing}
\label{app:Diff} 

In this appendix, we solve the generalized diffusion equations, \eqref{eq:DiffusionA}  for the quantum diffusion
modes in the presence of an external scalar dephasing field \eqref{dephasingpot}.

\subsection{Diffuson} 
\label{app:Diffuson}

We start out with the diffuson, for which
it is convenient to use central and relative times, $t=(t^+ +
t^-)/2$, $t'=t-\delt$ and $\eta = t^+-t^-$, such that 
\begin{align}
t^\pm = t\pm\eta/2,\qquad t^\pm-\delt = t'\pm\eta/2 .
\end{align}
In these variables, the differential equation for the diffuson 
$\Pi_\text{D}(\br,\br',t^+,t^-,\delt) = D^\eta_{tt'}(\br,\br')$
takes the form
\begin{align}
\label{appdiffusoneq}
\left( 
\partial_t - D \partial^2_\br 
- i   \phi_-^\eta(\br,t)
\right) 
D^\eta_{tt'}(\br,\br') 
&=  \delta(t-t') \delta(\br-\br'). 
\end{align} 
From here on we use the short notation  
\begin{align}
F_-^\eta(t)
&= F\left(t+\frac{\eta}{2}\right) - F\left(t-\frac{\eta}{2}\right)
\end{align} 
for arbitrary functions $F(t)$. For the classical diffuson one
has $\eta=0$ and thus the dephasing potential $\phi^0_- = 0$ disappears from the
problem, as it should. In the generalized diffuson however, particle and hole
visit the same position time-shifted by $\eta$, and dephasing occurs. 

Eq.~\eqref{appdiffusoneq} is equivalent 
to the imaginary-time Schr\"odinger equation for a particle of mass
$m=1/2D$ in  
a scalar potential $i\phi_-^\eta(\br,t)$. Its solution can be written as the path integral~\cite{Feynman, EfrosPollack} 
\begin{align}
\label{apppi}
D^\eta_{tt'}(\br,\br') 
&=  
\int_{\br(t')=\br'}^{\br(t)=\br} 
{\cal D}[\br(s)] \\
\times 
& \exp 
\left(
- \int_{t'}^{t}ds  \left\{ { \bold{\dot{r}}^2(s) \over 4D}
  + 
  i \phi_-^\eta(\br(s),s) \right\}   
\right).  \nonumber 
\end{align} 
We are interested in a potential that describes  momentum kicks via a
homogeneous force applied at well-defined instances
$t_1,t_2,...,t_N$ in time, 
$\hbar \phi(\br,t) 
= 
- \br\cdot\Delta\bold{p} f(t)$, 
where $\delp$  is the momentum transferred by
a single pulses, and  
$f(t)$ is a sum of functions peaked at the kick times $t_i$. We assume that the individual pulses are short
compared to their separation, such that $f(t)$ is zero
outside the vicinities $I_i$ of the $t_i$ and normalized to 
$\int_{I_i} dt f(t) = 1$. Aside this constraint, 
the following solution holds for arbitrary pulse shapes.  

To calculate the path integral 
we decompose the path $\br(s)$ connecting $\br'$ to $\br$
in the time $\delt=t-t'$ into a straight, ballistic trajectory plus fluctuations,  
$\br(s)= \overline{\br}(s)+ \widetilde{\br}(s)$. The ballistic path for $t' \leq s \leq t$ is 
$\overline{\br}(s) 
=\br'+ \frac{s-t'}{\delt}\Delta \br$, where $\Delta \br = \br -\br'$ and the closed loops 
from $\widetilde{\br}(t')=0$ to $\widetilde{\br}(t)=0$
can be written as the Fourier series 
$\widetilde{\br}(s) 
=
{1\over \sqrt{\delt}} \sum_{n\neq 0} \br_n e^{-i\omega_n s}$
 with 
$\omega_n = 2\pi n / \delt$.
Inserting into \eqref{apppi}, one notices that the two contributions decouple,
\begin{align}
D^\eta_{tt'}(\br,\br')
&=
\overline{D}^\eta_{tt'}(\br,\br')
\widetilde{D}^\eta_{tt'}. 
\end{align}
Only the ballistic contribution depends on the positions,
\begin{align}\label{diffDxxprime}
\overline{D}^\eta_{tt'}(\br,\br')
&=
\exp\left( 
-{\Delta\br^2\over 4 D\delt} 
+i  {\delp\over\hbar} \cdot \int_{t'}^t ds\, \overline{\br}(s) f_-^\eta(s)
\right), 
\end{align} 
where the $s$-independent components of $\overline{\br}(s)$ 
are weighted by the number 
\begin{align} 
\label{calFetadef} 
\chi_0^\eta(t',t) 
& =  \int_{t'}^t ds f_-^\eta(s).
\end{align} 
This is essentially the difference in the number of kicks experienced by particle and
hole during their evolution over the interval 
$[t'\pm\frac{\eta}{2}, t\pm\frac{\eta}{2}]$, respectively.  
For the classical
diffuson with $\eta=0$, these numbers are of course equal, and thus
$\chi^0_0(t',t)=0$. A priori, this need not be the case in the
general setting. If, then, particle and hole do not experience the
same number of kicks, this will result in uncompensated phases at all
times. 
Therefore, we
will consider in the following only those cases where  particle and
hole experience the same number of kicks (but possibly at different times), and
correspondingly make use of 
$\chi_0^\eta(t',t) = 0$. 

As a consequence, Eq.~\eqref{diffDxxprime} depends only on
the position difference, 
\begin{align}\label{diffDelx}
\overline{D}^\eta_{tt'}(\Delta\br)
&=
\exp\left( 
-{\Delta\br^2\over 4 D\delt} 
+i\frac{\chi_1^\eta(t',t)}{ \delt \hbar} 
 \delp \cdot \Delta \br
\right), 
\end{align} 
where the function 
\begin{align} 
\label{appchi1def} 
\chi_1^\eta(t',t) 
& =  \int_{t'}^t ds \, s\, 
f_-^\eta(s) 
\end{align} 
essentially evaluates the differences in particle and hole kick times.  
Fourier transformation in $\Delta\br$ then results in 
\begin{align}
\overline{D}^\eta_{tt'}(\bold{q})
&=\mathcal{N}^{-1}
\exp\left(-\frac{D}{\delt \hbar^2} \left[\delt\bq -\chi_1^\eta(t',t) 
 \delp \right]^2
\right),
\label{diffusonq}
\end{align} 
with normalization 
${\cal N}= (4\pi D\delt)^{-d/2}$. 

Concerning  the fluctuations, 
Gaussian integration over the $\br_n$ contributes the 
position-independent,
but  time-dependent contrast factor
\begin{align}
\label{appdiffgensol}
\widetilde{D}^\eta_{tt'} 
&= \mathcal{N}  
\exp \left[ 
-{D\delp^2\over\hbar^2} \left(
\chi_2^\eta(t',t) - {\chi_1^\eta(t',t)^2\over \delt}
\right)
\right], 
\end{align} 
where $\chi_1^\eta(t',t)$ of \eqref{appchi1def} appears squared, and 
\begin{align}
\label{appchi2def}
\chi_2^\eta(t',t) 
&=-\frac{1}{2}
 \int_{t'}^t  ds_1 
\int_{t'}^t  ds_2 \,
 |s_1 - s_2| 
f_-^{\eta}(s_1)  
f_-^{\eta}(s_2). 
\end{align}
When deriving the above expressions we have used 
that Gaussian integration over the $\br_n$ contributes the 
position-independent,
but  time-dependent contrast factor
\begin{align}
\label{app:matssum}
\widetilde{D}^\eta_{tt'} 
&= 
{\cal N} \exp \left[ 
- {D \delp^2 \over \delt \hbar^2} 
\sum_{n \neq 0} 
{ 
{\cal F}_n^\eta(t',t) {\cal F}_{-n}^\eta(t',t) 
\over 
\omega_n^2}  
\right],  
\end{align} 
where we introduced 
the pulse-difference Fourier transform 
${\cal F}_n^\eta(t',t) =
\int_{t'}^t ds\, 
e^{-i\omega_n s}
f_-^\eta(s)$.
The sum over frequencies in \eqref{app:matssum} 
is readily performed using that  
\begin{align}
\sum_{n\neq0} {e^{i\omega_n \Delta s}
\over \omega_n^2}
&=
{\delt^2\over 12} - \frac{\delt |\Delta s| }{2} + {\Delta s^2\over 2},
\label{Hcal}
\end{align}
with $\Delta s=s_1-s_2$, 
and upon employing $\chi_0^\eta(t',t) = 0$ 
(see discussion below eq.~\eqref{calFetadef}) 
one arrives at the stated result.

Summarizing we find the general diffuson
\begin{align} \label{appPiD}
\Pi_\text{D}(\bq,t^+,t^-,\delt) = e^{
-{D\over\hbar^2}  \left [
\delt \bq^2 - 2\chi_1(t^+,t^-,\delt) \bq\cdot\delp 
+\chi_2(t^+,t^-,\delt)\delp^2 \right]}, 
\end{align} 
where 
we returned to the time variables $t^\pm, T$ introduced in
the main text, and defined 
\begin{align} 
\chi_1 & (t^+,t^-,\delt) 
=\int_{-\delt}^0ds\,s\left[f(s+t^+) - f (s+t^-)\right] \nonumber \\
& = \int ds \, f(s) \left[(s-t^+)\chi_{\delt}^+(s) -
  (s-t^-)\chi_{\delt}^-(s) \right], 
\end{align} 
with 
$\chi_{\delt}^\pm(s):=\chi_{[-\delt,0]}(s-t^\pm)$ the characteristic function of the time
interval $[-\delt,0]$, evaluated for the particle at $s-t^+$ and the
hole at $s-t^-$. 
Similarly,  
\begin{align} 
\chi_2 & (t^+,t^-,\delt)  = -\frac{1}{2} \int ds_1  \int ds_2  f(s_1)f(s_2)
\nonumber \\ 
 & \times  \Big[ |s_1-s_2| \left\{
   \chi_{\delt}^+(s_1)\chi_{\delt}^+(s_2) +
   \chi_{\delt}^-(s_1)\chi_{\delt}^-(s_2)\right\} 
 \nonumber \\ 
& \qquad - 2  |s_1-t^+-s_2+t^-| \chi_{\delt}^+(s_1) \chi_{\delt}^-(s_2)\Big].
\end{align} 
Specialized to $\delta$-pulses $t_1$ and $t_2$
the above expressions turn into eqs.~\eqref{appchi1d} and following used in section~\ref{app:D2echo}.

\subsection{Cooperon} 
\label{app:Cooperon}

Turning to the Cooperon, it is again convenient to use central and relative times, $t=(t^+ +
t^-)/2$  and $\eta = t^+-t^-$, as well as $\eta-\eta' = \deleta = 2\delt$, such that 
\begin{align}
t^\pm = t\pm\eta/2,\qquad t^\pm\mp \delt = t\pm\eta'/2 .
\end{align}
In these variables, the Cooperon differential equation for
$\Pi_\text{C}(\br,\br',t^+,t^-,\delt) = C^t_{\eta\eta'}(\br,\br')$
takes the form
\begin{align}
\label{appCooperoneq}
\left( 
\partial_\eta - \frac{D}{2} \partial^2_\br 
- \frac{i}{2}   \phi_-^\eta(\br,t)
\right) 
C^t_{\eta\eta'}(\br,\br') 
&=  \delta(\eta-\eta') \delta(\br-\br'), 
\end{align} 
where $ \phi_-^\eta(t) = \phi(t+\frac{\eta}{2}) -
\phi(t-\frac{\eta}{2})$ as before. For the single-mode Cooperon, equality
of starting and end times imposes $\eta'=-\eta$. 
In difference to the diffuson case, the dephasing potential stays in the problem, 
and we now have to solve the equation of motion in $\eta$ at
fixed $t$. This is achieved with the path integral~\cite{Feynman, EfrosPollack}
\begin{align}
\label{apppiC}
C^t_{\eta\eta'}(\br,\br') 
&=  
\int_{\br(\eta')=\br'}^{\br(\eta)=\br} 
{\cal D}[\br(u)] \\
\times 
& \exp 
\left(
- \int_{\eta'}^{\eta}du  \left\{ { \bold{\dot{r}}^2(u) \over 2D}
  + 
  \frac{i}{2} \phi_-^u(\br(u),t) \right\}   
\right).  \nonumber 
\end{align} 
Following then the same steps as before for the diffuson one arrives at the
dephased general Cooperon (expressed in time variables $t^\pm, T$ used in
the main text)  
\begin{align} 
\Pi_\text{C}(\bq,t^+,t^-,\delt) & =e^{
-{D\over\hbar^2}  \left [
\delt \bq^2 - 2\bar\chi_1(t^+,t^-,\delt) \bq\cdot\delp 
+\bar\chi_2(t^+,t^-,\delt)\delp^2 \right]}, 
\label{PiCqt}
\end{align} 
where 
\begin{equation} 
\bar\chi_1  (t^+,t^-,\delt) 
 =  \int du\, 
 f(u) \left[ (u-t^+)\bar\chi_{\delt}^+(u) +(u-t^-)\bar\chi_{\delt}^-(u)\right] ,
 \label{barchi1} 
 \end{equation} 
with
$\bar\chi_{\delt}^\pm(u):=\chi_{[-\delt,0]}(\pm(u-t^\pm))$ the characteristic function of the time
interval $[-\delt,0]$, evaluated for the particle at $u-t^+$ and the
hole at $t^--u$. Similarly,  
\begin{align} 
\bar\chi_2 & (t^+,t^-,\delt)  =  - \frac{1}{2} \int du_1  \int du_2  f(u_1)f(u_2)
\nonumber \\ 
 & \times \Big[ |u_1-u_2| \left\{
   \bar\chi_{\delt}^+(u_1)\bar\chi_{\delt}^+(u_2) +
   \bar\chi_{\delt}^-(u_1)\bar\chi_{\delt}^-(u_2)\right\} 
 \nonumber \\ 
& \quad - 2  |u_1+u_2-t^+-t^-| \bar\chi_{\delt}^+(u_1) \bar\chi_{\delt}^-(u_2)\Big].
\label{barchi2} 
\end{align} 

The single-mode Cooperon evaluated at 
$t^+= t = \delt $ and $t^-=0$ then reads 
\begin{align} 
\Pi_\text{C}(\bq,t) & =e^{
-{D\over\hbar^2}  \left [
t \bq^2 - 2\bar\chi_1(t) \bq\cdot\delp 
+\bar\chi_2(t)\delp^2 \right]}, 
\label{PiCqtoneloop}
\end{align} 
where the auxiliary functions in the exponential are 
\begin{align}
\bar\chi_1(t)&= \int_0^t du (2u-t) f(u), \label{bchi1}
\\
\bar\chi_2(t)&=  \int_0^t du \int_0^t dv f(u) f(v) 
(|u+v-t| - |u-v|).   \label{bchi2}
\end{align}
For a single $\delta$-pulse $f(t) = \delta(t-t_1)$, these functions
become 
$\bar\chi_1(t) = (2t_1 - t)\theta(t-t_1)$, and         
$\bar\chi_2(t)  = |2t_1-t|\theta(t-t_1)$ 
as used in section~\ref{MSE}.
From the general expressions \eqref{bchi1} and \eqref{bchi2} 
we further find the features also discussed there, i.e. 
for a pulse of finite resolution in time but still symmetric around
$t_1$, $\int dt \, t \, f(t) =  \langle
t \rangle_f =  t_1$  defines the dephasing pulse center, and  
thus $\bar\chi_1(2 t_1) = 0$ by construction.
 At this instant, the entire 
contrast penalty vanishes, since $\bar\chi_2(t)$ vanishes by
symmetry as well, implying a perfect revival. Only for an asymmetric pulse the
contrast penalty will generically remain finite,
since then $\bar\chi_2(t)$  is not required to vanish
exactly at $2t_1$, and the echo will appear with reduced contrast. 
Finally, for a sequence of two $\delta$-kicks 
\begin{align}
\bar\chi_1(t) & = 2(t_1+t_2 - t), \\        
\bar\chi_2(t) & = |2t_1-t|  +  |2t_2-t| + 2 |t_1+t_2-t|  - 2 |t_2-t_1|,  
\end{align}
which results in a C1 echo at time $\tau_2=t_1+t_2$.

\end{appendix}

\end{document}